\documentclass{ws-ijmpe}
\begin{document} 
\markboth{Rudolph C. Hwa}{Some Closing Remarks}

\title{CORRELATION AND FLUCTUATION IN MULTIPARTICLE PRODUCTION:  SOME CLOSING REMARKS}

\author{RUDOLPH C. HWA}

\address
{Institute of Theoretical Science and Department of
Physics\\ University of Oregon, Eugene, OR 97403-5203, USA}

\maketitle

\begin{abstract}
Some general comments are made on the evolution of this series of workshops and on some features of this particular Workshop without attempting to summarize all the talks presented.
\end{abstract}
\vspace*{.5cm}
I was asked to close this Workshop with some comments that can give a historical perspective of the series.  I will therefore not attempt to summarize the talks that have been presented here.  This series of workshops have been held over the span of about twenty years, and I happen to be the only one who has been to all eleven of them.  The titles and contents of the workshops have varied over the years, but the invariant theme has always been about multiparticle production.

The first workshop held in Aspen in 1986 was a very small one and did not have the format that has been customary for all others in the series.  There were no proceedings to look back to.  The first formal one was held in Jinan, China in the following year, organized by Q.-B.\ Xie and me.  In the prologue to the proceedings I wrote:  ``Multiparticle production is regarded by particle physicists as being too complicated and therefore declared uninteresting.  $\cdots$ The malaise is basically an internal one.  The absence of a workable theory gave rise to a proliferation of models, many of which bear little resemblance to one another. $\cdots$ The field not only has no first principles to rely on, $\cdots$ it has generated no collective wisdom to serve as a guide to the construction of models.''  In the International Conference on High Energy Physics (Berkeley, 1986) there was not one session devoted exclusively to soft hadronic processes and multiparticle production.  The need for a workshop was obvious.

The situation today is totally reversed in the nuclear physics community, where the complexity of a collisional system does not pose a barrier to the motivated investigators, and where the applicability of relativistic hydrodynamics to heavy-ion collisions provides a successful theoretical basis for the description of the evolutionary properties of the bulk matter, in essentially the same framework formulated by Landau\cite{ll} and Bjorken\cite{jb} for hadronic collisions.  The issue is no longer how to describe the collision process, but what we can learn from the hydrodynamical parameters and how to understand the features that cannot be explained by hydrodynamics.  This Workshop on correlation and fluctuation is aimed at addressing precisely those issues.

Another way to see the difference between 1987 and 2006 is that at Jinan there were only four major experimental talks (not counting two very minor such talks), i.e., from NA22, NA9 (EMC) \& WA21, UA1 and UA5, whereas at Hangzhou there were 19 experimental talks, all except one being on RHIC data.  Even those 18 talks represent only a small sample of the mountain of data from RHIC.  As the dictum goes, physics is driven by experiment.  In 1987 negative binomials were tested to fit multiplicity distributions; today they are used to parameterize event-by-event fluctuations.   Jet physics was getting started in the 80s; today it is used for tomography of dense medium.  Without the RHIC data the knowledge gained from the past was not used in tools to facilitate the exploration of new problems, which is why multiparticle production lacked the stimulus to bring it to the frontier of high-energy physics toward the end of the last decade.  RHIC changed all that.  It may be nuclear physics, but it is at high energy.  Interesting physics emerges from the interplay between perturbative and non-perturbative parts of high-energy nuclear physics, giving new life to the subject of multiparticle production.

Jet physics studied intensively in leptonic and hadronic collisions is, however, only one of many areas having high level of activities at RHIC.  Bose-Einstein (BE) correlation and event-by-event (EbE) fluctuation have also been subjects of particular interest at RHIC, not because of applicability of perturbative methods, but because space-time property and critical behavior are of primary importance in large and dense systems that cannot be achieved in collisions of elementary particles.  These three topics, {\it viz}, (a) BE correlation, (b) EbE fluctuation, and (c) correlation in jets, are therefore the subjects addressed by the main sessions of this Workshop.

Over the years the emphases in past workshops have shifted from one subject to another.  In the early 90s intermittency was a hot topic.  That led in the mid-90s to BE correlations and critical fluctuation.  Jets have always been discussed in big meetings, and were included in this series of workshops mainly in the context of self-similarity.    The three topics of this Workshop are therefore very natural extensions of the ongoing concerns in multiparticle production.  Nevertheless, it cannot be denied that researchers in the different areas have denominational interests that do not overlap.  In a large conference, such as Quark Matter, talks on the three subjects may be presented in sessions that run in parallel without encountering too many objections arising from participatory conflicts.  So what justification do we have in a small Workshop like this to include all three areas?

In my attempt to answer this question, as well as to give an overall view of what has been discussed here, let me venture to use a metaphor:  how dragons have been depicted in different cultures in very different ways.  In China it represents good fortune, wisdom and longevity, and in gold color it represented the emperor with its imperial power.  In England it is treacherous and demonic and must be killed; it took the courage of St.\ George to do it.  In the Germanic operas {\it Der Ring des Nibelungen} by R.\ Wagner the dragon in {\it Siegfried} is a lazy and greedy beast that only wants to be left alone to guard the golden treasure it has hoarded.  The point relevant to us is that Siegfried, the hero, kills the dragon with Nothung (his sword), tastes the blood of the dragon, and gains knowledge.

\begin{figure}[th]
\hspace*{-.6cm}{\psfig{file=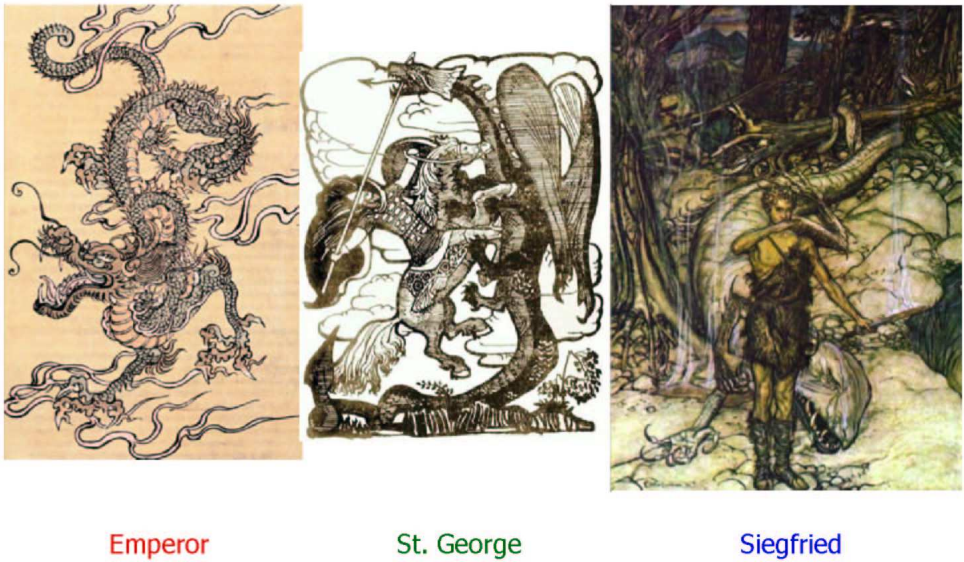,width=14cm}}
\vspace*{-3cm}
\caption{Dragons as metaphor}
\end{figure}

The dragon is like the quark-gluon plasma (QGP):  in both cases they are hard to find, and if found, we would like to know what they really are.  The ways in which the dragon is regarded in the different cultures illustrate the different ways that the QGP is treated in the three sessions of this Workshop.

BE correlation studies the QGP from far away, like the Chinese dragon.  The space-time properties of the collisional system are revealed in the correlation of nearby momentum vectors of identical particles emitted from the system.  The dragon is not disturbed by spear or sword, but the image it projects is sufficient to generate a HBT puzzle that has ruled out many models.  The HBT analysis has the virtue of being able to constrain the description of the space-time evolution of the medium generated in heavy-ion collisions.  The survivors of the HBT analysis are such models as AMPT, hadronic cascade, Buda-Lund, Cracow, blast-wave, Renk, granular QGP, etc.\cite{tc}  They still represent a wide variety of evolutionary systems but at least they can reproduce the $k_T$ dependences of the three radii, $R_{\rm side}$, $R_{\rm out}$ and $R_{\rm long}$ that have been observed.

Very few phenomenological papers on BE correlation were presented at this Workshop.  Theoretical papers presented by Danielewicz,\cite{pd} Enokizono,\cite{ae} Eggers,\cite{he} and Mrowczynski\cite{sm} are all aimed at improving the analysis. The general recognition is that the emission sources are not Gaussian.  The analyses of the shape using Cartesian and spherical harmonics have been a major effort to reconstruct 3D source function without getting involved with the dynamics, a good example of observing the dragon from afar.  Wong and Zhang, on the other hand, conjecture that large density fluctuations may be modeled by granular droplet sources, which in turn may account for the HBT radii, $p_T$ spectra and elliptic flow.\cite{cw,wz}  I am doubtful that the $p_T > 2$ GeV/c region can be successfully included in this framework without considering hard scattering of partons.  Nevertheless, they have moved closer to the dragon and proposed some gross space-time properties that it may possess.

To get more information about the mythological beast, we may have to get close enough and kill it.  Can the life-death transition tell us about a phase transition of the first and second order?  We may have to find many dragons and take more than one St.\ George to slay them to observe the EbE fluctuation of how the dragons die, thereby learning about their critical behavior.  Many measures have been proposed to quantify such fluctuations.  T.\ Nayak has given a very broad overview of the subject.\cite{tn}  While a good summary of the data in a wide variety of measures is presented, very little seems to pertain to the issue of critical behavior, which is the {\it raison d'$\hat{e}$tre} for the investigation.  Mitchell comes closest to the issue by showing the correlation length determined from the PHENIX data.\cite{jm}  He first shows the multiplicity fluctuations in terms of the negative binomial parameter $k$, whose dependence on the rapidity interval $\delta \eta$ enables the extraction of  the correlation length $\xi$.  It is found that $\xi$ exhibits universal scaling and decreases with $N_{\rm part}$.  But the values of $\xi$ are so small at any $N_{\rm part}$ that there is no hint of any critical behavior, for which $\xi$ is supposed to diverge at the critical point in an infinite system.

The problem, as I see it, is that critical exponents have been studied exhaustively for static systems, whereas QGP is an evolving system with thermodynamical quantities depending on spatial coordinates and time.  Thus in each event the system undergoes phase transition, if it does, at different points in space at different times.  All hadrons produced in each event go to the detector and are superimposed in the event structure that has no way to discriminate times of creation. How can correlations that exist at one and the same time of emission be disentangled from those that exist at later times?  EbE fluctuations take each event as a package in its whole, but different parts of a dragon may die at different times.

Finally, we come to the biological structure of the dragon.  Since we have never found any archeological remains of the mythological beasts, they must vanish after being slain, like Obi-Wan Kenobi in {\it Star Wars}.  Some MRI records taken while they are still alive would be useful. That is where jet tomography comes in.  My feeling is that due to the narrow path of a hard parton trajectory in the dense medium the utility of jet tomography is limited, unlike an X-ray picture that exhibits the whole 2D projection at once.  Nevertheless, the study of hadron correlations in jets has been very fruitful.  At this Workshop we have seen the most intense interaction between theory and experiment in that subject.  The jet is like Nothung, Siegfried's sword, that pierces the dragon's heart.  In drawing the sword Siegfried gets some of the dragon's blood on his hands, which he instinctively put to his mouth to suck.  By tasting the blood Siegfried is empowered to understand the meanings of  the songs of the forest bird.  For us the blood is like the associated particles in a jet that empower us to gain knowledge on what we seek.

M.\ van Leeuwen and J.\ Jia have given excellent reviews of correlations in jets, as observed by STAR and PHENIX, respectively.\cite{ml,jj}  Among the topics covered,  a very active area of investigation in the past year or two, both experimentally and theoretically, has been the $\Delta \phi$ distribution on the side away from the trigger in the transverse plane.  The dip-bump structure observed around $\Delta \phi = \pi$ has given rise to several theoretical explanations, which in turn have stimulated further analyses to discriminate and quantify the magnitudes of different possible contributions to the effect.  Most of the talks on the first day of this Workshop address that problem, and need not be mentioned individually here.  An exciting possibility is that the observed $\Delta \phi$ distribution is a manifestation of the collective response of the medium to the passage of a hard parton, such as the development of shock wave.  Experimentally, the magnitude of that effect seems to be weaker than that due to deflected jets.\cite{fw}  Furthermore, the dip-bump structure seems to exist even at very large rapidities $(2.7 < |\eta| <3.9)$ in recoil against a trigger at mid-rapidity.\cite{fw2}  Since the local densities at large $|\eta|$ are lower than that at $\eta \sim 0$, one would expect the properties of the shock wave, if excited, to be very different from what has been observed in the $\Delta\phi$ distribution.\cite{fw2}  Whatever final interpretation of the effect can be agreed upon eventually, there is no question that the physics involved is just what jet tomography is supposed to probe.

The study of jets is likely to become very different at LHC, where jet production will not be rare, unless $p_T$ is extremely large.  For $p_T < 20$ GeV/c, so many jets will be produced that the correlations we have examined at RHIC are meaningless in the presence of a sea of associated particles all contributing to the statistical background.  If the trigger momentum is set very high, such as $>50$ GeV/c, then the interaction of the hard parton with the medium will be very different from what we have come to be familiar with, which is most effective when the parton momentum is moderate and the medium effect is strongest.  Perhaps we shall have to find another culture in which the dragon is perceived in a way different from what we have mentioned here.

Instead of speculating on what that dragon might be, let us appreciate what we have experienced in the last four days, and thank the organizers for this successful and enjoyable Workshop and for sighting a dragon boat on the West Lake.

This work was supported,  in part,  by the U.\ S.\ Department of Energy under Grant No. DE-FG02-96ER40972.

\end{document}